\documentclass[a4paper,10pt,twoside]{cpc-hepnp}

\usepackage{multicol}
\usepackage{graphicx}
\usepackage{booktabs}
\usepackage{amssymb,bm,mathrsfs,bbm,amscd}
\usepackage[tbtags]{amsmath}
\usepackage{lastpage}
\usepackage{CJK}
\usepackage{textcomp}

\begin{document}
\begin{CJK*}{GBK}{song}

\title{Measurement of photoelectron yield of the prototype for the CDEX-10 liquid argon detector}

\author{%
      CHEN Qing-Hao$^{1,2;1)}$\email{cqh10@mails.tsinghua.edu.cn}%
\quad YUE Qian$^{1,2}$
\quad CHENG Jian-Ping$^{1,2}$\\
      KANG Ke-Jun$^{1,2}$
\quad LI Yuan-Jing$^{1,2}$
\quad LIN Shin-Ted$^{3}$
\quad TANG Chang-Jian$^{3}$\\
      XING Hao-Yang$^{3;2)}$\email{xhy@scu.edu.cn}
\quad YU Xun-Zhen$^{3}$
\quad ZENG Ming$^{1,2}$
\quad ZHU Jing-Jun$^{3}$
}
\maketitle

\address{%
$^1$ Department of Engineering Physics, Tsinghua University, Beijing 100084, China\\
$^2$ Key Laboratory of Particle and Radiation Imaging, Tsinghua University, Ministry of Education, China\\
$^3$ College of Physical Science and Technology, Sichuan University, Chengdu 610064, China\\
}

\begin{abstract}
The China Dark Matter Experiment (CDEX) is a low background experiment at China Jinping Underground
Laboratory (CJPL) designed to directly detect dark matter with a high-purity Germanium (HPGe)
detector. In the second phase CDEX-10 with a 10 kg Germanium array detector system, the liquid argon (LAr)
anti-compton active shielding and cooling system is proposed. For purpose of studying the properties of
LAr detector, a prototype with an active volume of 7 liters of liquid argon was built and operated. The
photoelectron yields, as a critically important parameter for the prototype detector, have been measured to be
0.051-0.079 p.e./keV for 662 keV Gamma lines at different positions. The good agreement between the
experimental and simulation results has provided a quite reasonable understanding and determination of the
important parameters such as the Surviving Fraction of the Ar$^*_2$ excimers, the absorption length for 128 nm photons
in liquid argon, the reflectivity of Teflon and so on.
\end{abstract}



\begin{multicols}{2}

\section{Introduction}

Weakly Interacting Massive Particles (WIMPs) are the candidates of dark matter in our universe.
Up to now any direct interaction of WIMPs with nuclei has not been observed yet. Since 2009, the
CDEX Collaboration has been working on searching for low mass WIMPs with HPGe
detector\cite{CDEX_introduction1,CDEX_introduction2}. Several results have
been produced by the first phase CDEX-1\cite{CDEX_result1,CDEX_result2,CDEX_result3}. In
the second phase CDEX-10, the HPGe detector is designed to be immersed in liquid argon which
serves as an anti-Compton detector. In this design, the LAr veto detector also serves as both
the passive shielding detector and the low temperature medium for the HPGe detector.

Liquefied noble gases as detection medium for dark matter are widely used by a number of existing and
proposed experiments\cite{XENON,Darkside}. Liquid argon has high scintillation yield ($\sim$40 photons/keV\cite{scinYield})
and is easily purified and scalable to large mass with relatively low cost, which makes it promising
for this application.

The sensitivity of the LAr veto detector is strongly dependent on the number of detected
scintillation photons. In order to develop and optimize the light collection of the
liquid-argon-based detector, a prototype detector was designed and manufactured in Tsinghua
University and then tested in Sichuan University.

In the following sections we will describe the prototype, the single-photoelectron calibration,
and we report the photoelectron yield measured with $^{137}$Cs 662 keV lines at different
collimated positions.

\section{The prototype of the liquid argon detector}

In this section, we will describe the device in detail, especially the light collection system
of the prototype detector.

\subsection{The detector system}

Fig.~\ref{fig1} shows the schematic diagram of the LAr detector system. The detector system is
composed of the following four parts: LAr Dewar, purification system, cooling system, vacuum pump.

The LAr Dewar is the main part of the prototype detector system. As shown in Fig.~\ref{fig2}, the
vacuum-insulated stainless steel Dewar contains a 7-liter active volume of LAr viewed by a
206-mm-diameter 8" ETL 9357 photomultiplier tube (PMT) with a quantum efficiency (QE)
of $\sim$17\%\cite{PMT}. The active LAr region is defined by a Teflon
cylinder which is 150 mm in diameter and 400 mm high. The PMT is held in place by two Teflon rings
below and above. To detect the 128 nm argon scintillation light, we
\begin{center}
\includegraphics[width=7cm]{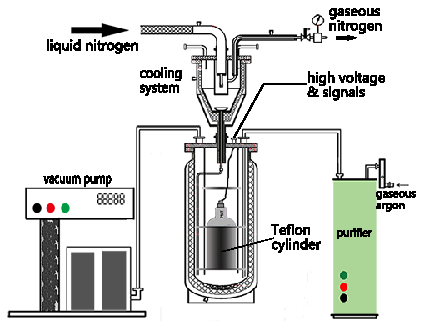}
\figcaption{\label{fig1}   Schematic diagram of the prototype detector system.}
\end{center}
use the tetraphenyl butadiene (TPB) as the wavelength
shifter (WLS), with a peak emission wavelength of 420 nm\cite{lab1}. TPB is deposited by
vacuum evaporation onto the inner surface of the Teflon cylinder and the PMT window (see section 4.2 for
more discussion) with the thickness of $\sim$0.3 mg/cm$^2$. In addition, by placing a piece of stainless
steel above the PMT, the heat convection and the evaporation rate can be reduced.
\begin{center}
\includegraphics[width=5.5cm]{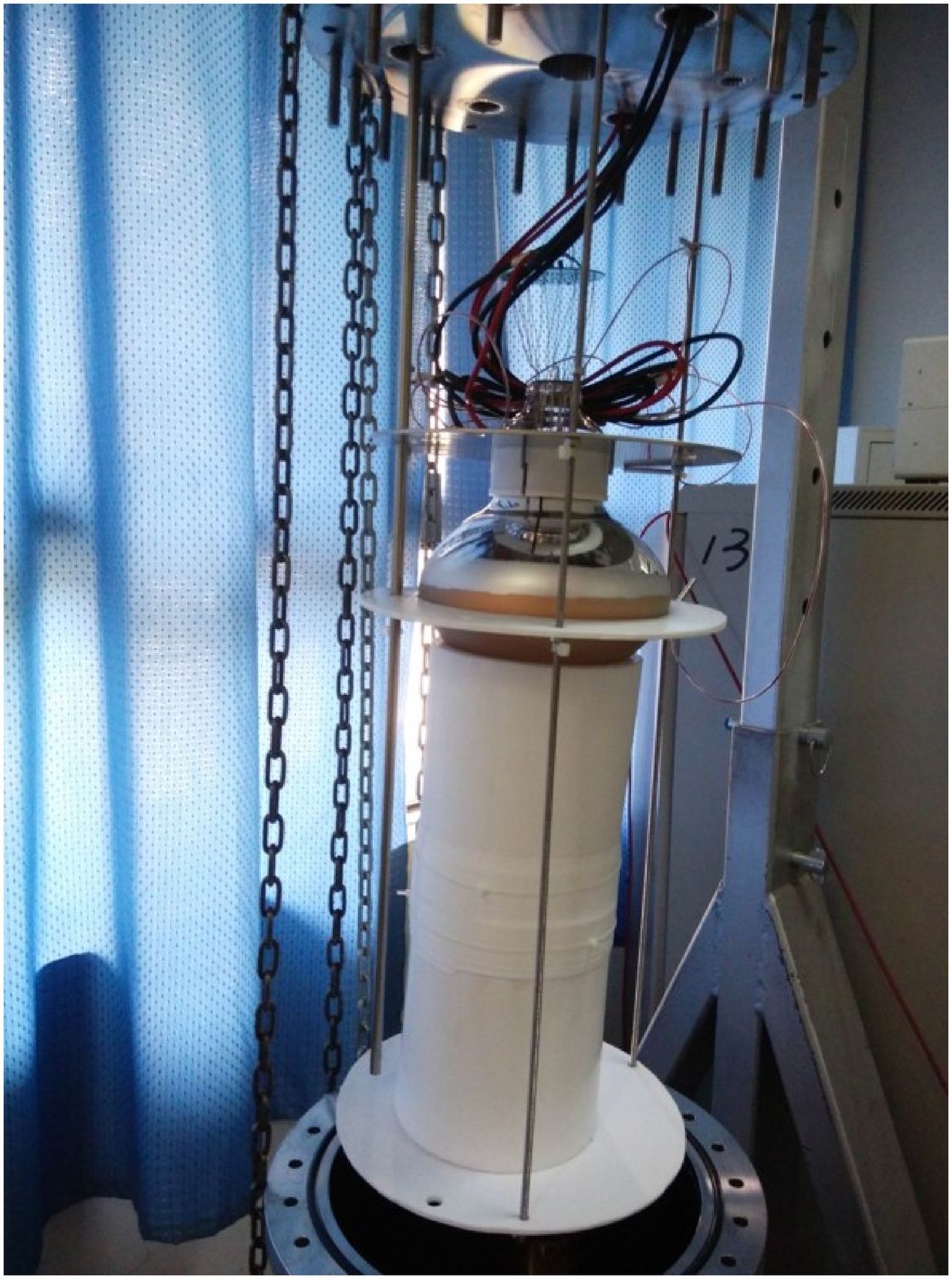}
\figcaption{\label{fig2}   The layout of LAr container and PMT.}
\end{center}

The LAr Dewar is evacuated by the vacuum pump so that the gas impurities can be removed before running the
detector. The cooling system, composed of a cooling chamber and liquid nitrogen supplied from outside,
is employed to maintain the temperature of LAr during run time. Argon gas enters the cooling chamber and is
then liquefied by the heat exchanger (made of copper). Liquefied fluid flow back into the Dewar through a pipe.

Previous studies indicate that the contamination in LAr can seriously reduce the scintillation light intensity
through collisional de-excitation of Ar$^*_2$ excimers\cite{lab2}.
The LAr used in the experiment was commercially bought and the contamination level is lower than 10 ppm.
Impurities in the detector can build up over time via outgassing, and reduced LAr purity.
Hence the purification system is proposed in the next stage of experiment. A SAES getter will be used to
ensure high purity of LAr.

More details about the detector structure and the TPB coating process can be found in Ref.\cite{previousPaper}.

\subsection{The data acquisition system}

Each anode signal is fan out as two identical signals: one is fed into the discriminator to provide
the trigger signal and the other is sent to a 8 bit, 50 MS/s digitizer (CAEN V1721) to digitize the signal
pulse. The data are saved into hard disk for offline analysis.
The trigger threshold is set to about 1/2 of the single photoelectron mean amplitude. To
measure the DAQ dead time, the discriminator output of the
random trigger events (10 Hz) from a pulse generator served as a trigger and were digitized
as well. When the digitizer is triggered by an event, the data stream is then downloaded
and stored on a local hard disk. The time window of each event is 20 \textmu s (5 \textmu s before the
trigger and 15 \textmu s after).

\section{Single-photoelectron calibration}

For a measurement of the photoelectron yield of the prototype detector, we need a gain calibration of
the PMT. An LED calibration procedure was used to evaluate the charge response of the PMT to single
photoelectron. The PMT was immersed in liquid argon and the light pulses of $\sim$10 ns duration
at $\sim$440 nm wavelength from an LED were injected in through an optical fiber that
terminates on the PMT window. The intensity of the blue light can be adjusted by setting
the voltage of the LED.

The number of photoelectrons (p.e.) induced by each incident light pulse follows
a Poisson distribution and the probability of having $n$ photoelectrons is expressed as\cite{lab3}:
\begin{equation}
\label{eq1}
P(n)=\mu^n\cdot\frac{{\rm e}^{-\mu}}{n!}
\end{equation}
Where the mean value $\mu$ is determined by the light intensity. To evaluate the charge response
of the PMT to a single photoelectron, it is necessary to ensure
that most of the signals come from single electron events, which can be achieved by requiring
the number of signals with two photoelectrons being below, for instance, 10\% of that of single
photoelectron. According to Eq.~(\ref{eq1}), we can write:
\begin{equation}
\label{eq2}
\frac{P(2)}{P(1)}=\frac{\mu}{2}=0.1
\end{equation}
Eq.~(\ref{eq2}) can be translated into $\mu$ = 0.2 and $P(0)$ = 81.9\%. Hence, the number of 2
photoelectrons signals will be 10\% lower than that of 1 photoelectron if the light
intensity is adjusted such that the number of empty triggers is 81.9\%. Under this condition,
for any other number of photoelectrons the probability will be negligible.

The PMT response to photoelectrons can be roughly treated as a series of Gaussian functions
of different weight. Taking account of the electronic noise, an exponential functions should
be added to the PMT response function\cite{lab4}. For our purpose, however, a simplified fit
function $f(x)$ is sufficient:
\begin{eqnarray}
\label{eq3}
f(x)=a\cdot{\rm e}^{-bx}+\sum^3_{n=1}c_n\cdot G(n\mu_1,\sqrt{n}\sigma_1)
\end{eqnarray}
The exponential function with fit parameters $a$ and $b$ represent the pedestal which is
dominated by electronic noise. The Gaussian functions $G(n\mu_1,\sqrt{n}\sigma_1)$, with mean value
$n\mu_1$, standard deviation $\sqrt{n}\sigma_1$, and a normalization constant $c_n$, are the sum of
the single photoelectron response ($n$=1) and multi-photoelectron responses ($n>$1).

A function generator pulses the LED at a certain frequency and triggers the DAQ system simultaneously.
Each signal of the PMT output was collected. After subtraction of the baseline, the integral of the recorded
waveform is evaluated within a fixed 100-ns window around the arrival time of the light pulse. The charge
spectrum of the PMT was then obtained and fitted with Eq.~(\ref{eq3}), as shown in Fig.~\ref{fig3}.
The parameter $\mu_1$, with a value of 4.625 pC, is the absolute gain of the PMT for a single photoelectron. More
test on the performance of the PMT operating at both room and cryogenic temperatures can be
found in Ref.\cite{dys}.
\begin{center}
\includegraphics[width=9cm]{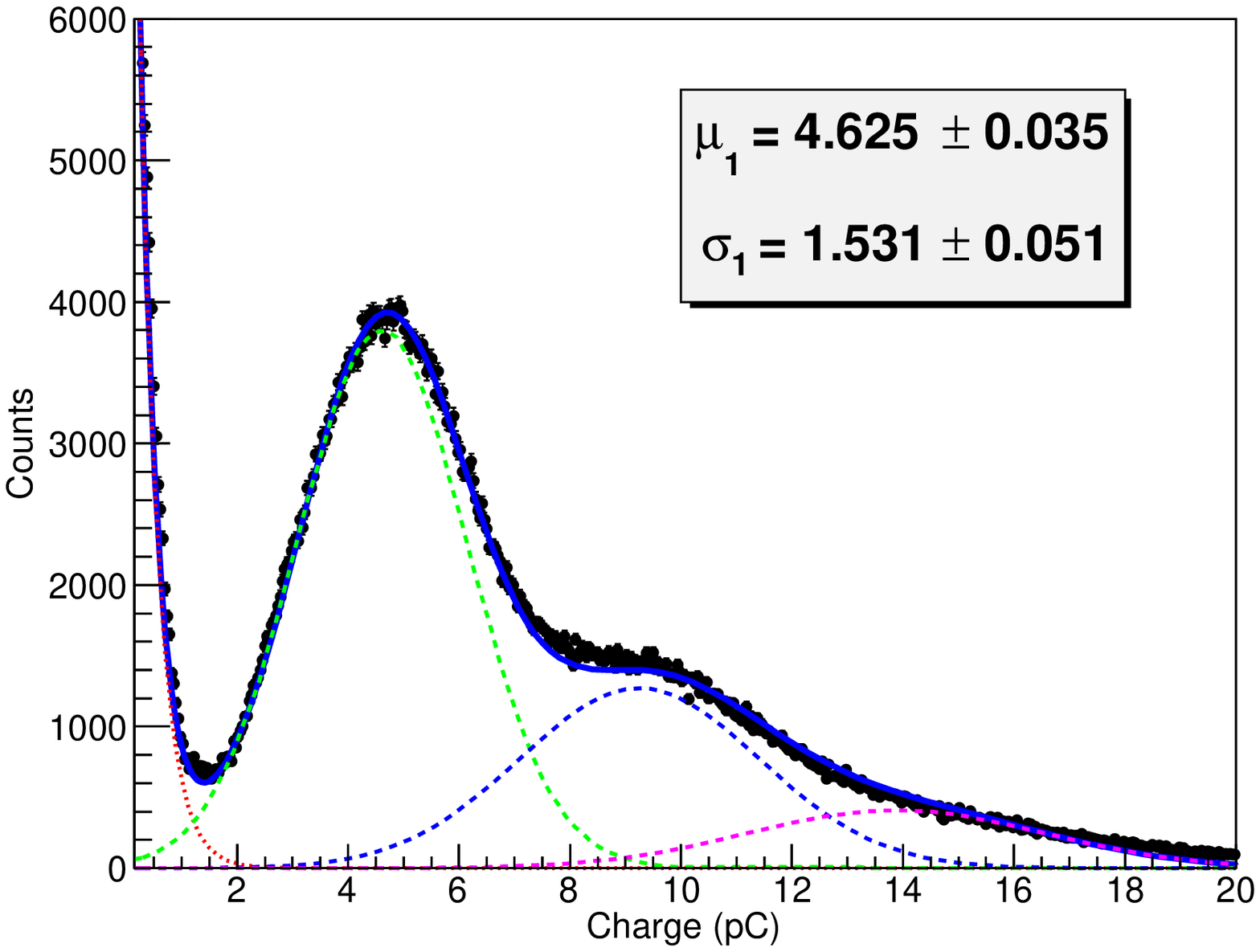}
\figcaption{\label{fig3}   The single photoelectron response spectrum of the ETL 9357FLB PMT at
1600V at liquid argon temperature. The dotted red curve is the exponential part of the fit function. The three
dashed curves (green, blue, pink) are the gaussian parts of the fit function and represent 1-3 p.e. response
respectively.}
\end{center}

\section{The Gamma ray response of the prototype detector}

In this section, the photoelectron yield of the prototype detector is estimated by exposing the
detector to an external $\gamma$ source collimated at different positions. The difference in
photoelectron yield between coating the PMT window or not is figured out.

The Gamma source we used is $^{137}$Cs with an intensity of 10 mCi. The Gamma lines are collimated
by a 50-mm-thick lead collimator with a 10 mm diameter hole. Three positions of collimated Gamma source along the
vertical direction are used: A, B and C, corresponding to 100 mm, 200 mm and 300 mm from the bottom
of the active volume (shown in Fig.~\ref{position}).
\begin{center}
\includegraphics[width=3.1cm]{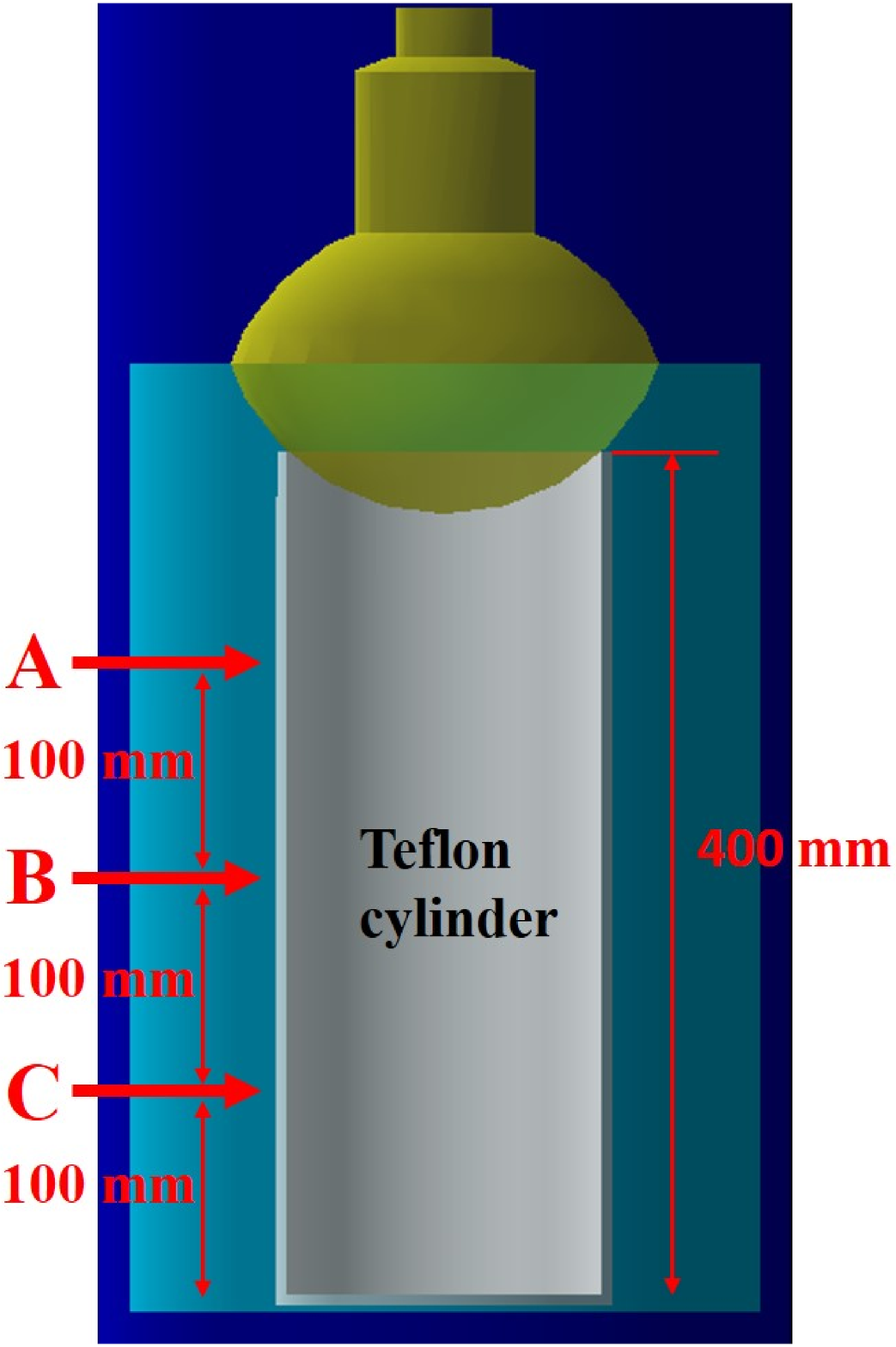}
\figcaption{\label{position}    Schematic diagram of the three positions of collimated Gamma source.}
\end{center}

\subsection{The photoelectron yield for 662 keV Gamma lines}

The Gamma-induced scintillation spectra for the three positions of collimated Gamma source, shown in Fig.~\ref{fig4}-\ref{fig6},
were obtained after subtraction of a background spectrum acquired with no source present. The two peaks
in each spectrum are the compton edge (left) and the full absorption peak (right), respectively. They
are consistent quite well with the spectra from a GEANT4-based Monte Carlo simulation.
\begin{center}
\includegraphics[width=8.5cm]{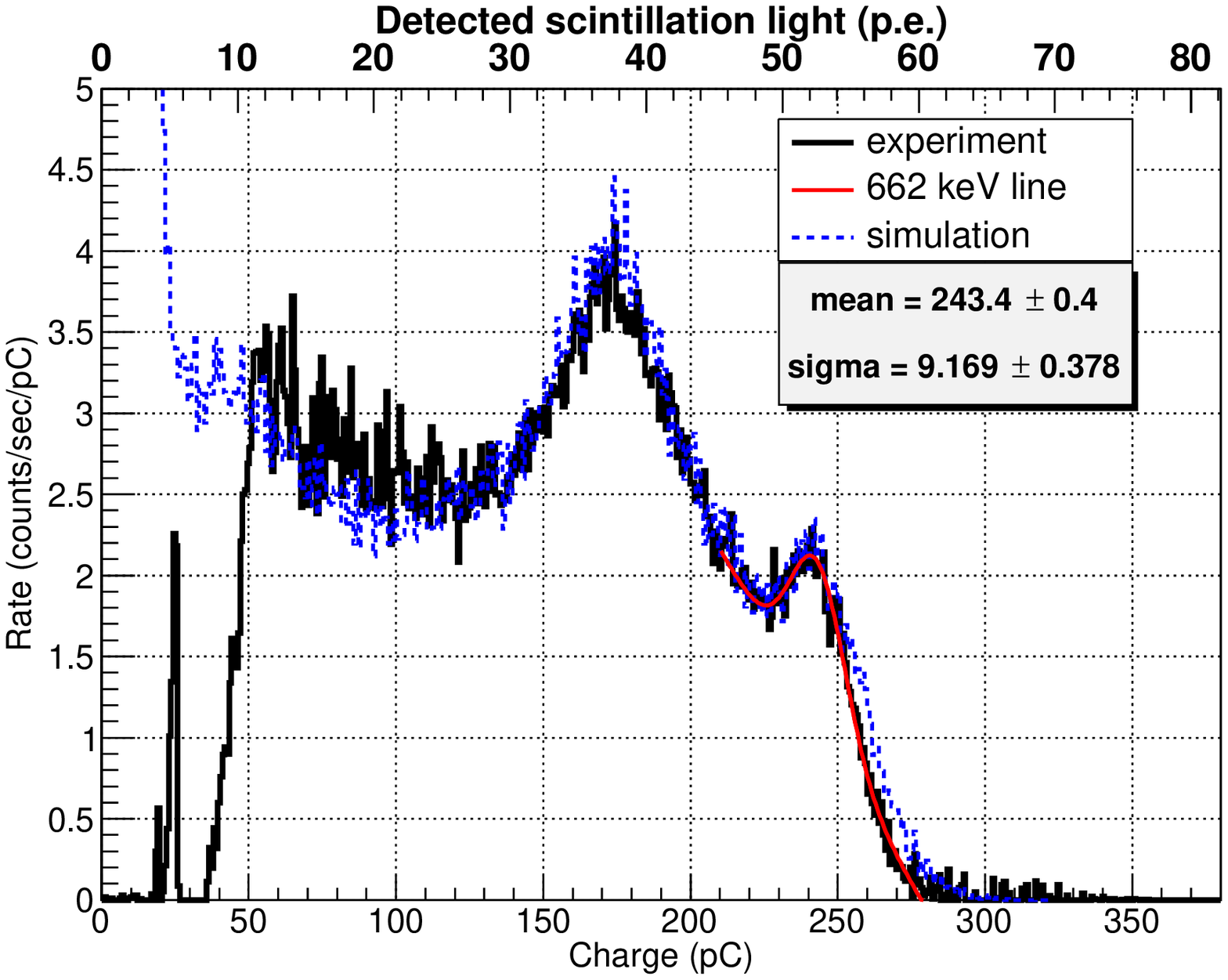}
\figcaption{\label{fig4}   Scintillation spectrum of $^{137}$Cs collimated at position A (solid black
histogram), as well as the Monte Carlo simulation result (dashed blue histogram). The full absorption
peak is fitted with the sum of a Gaussian and a linear function(red line).}
\end{center}
\begin{center}
\includegraphics[width=8.5cm]{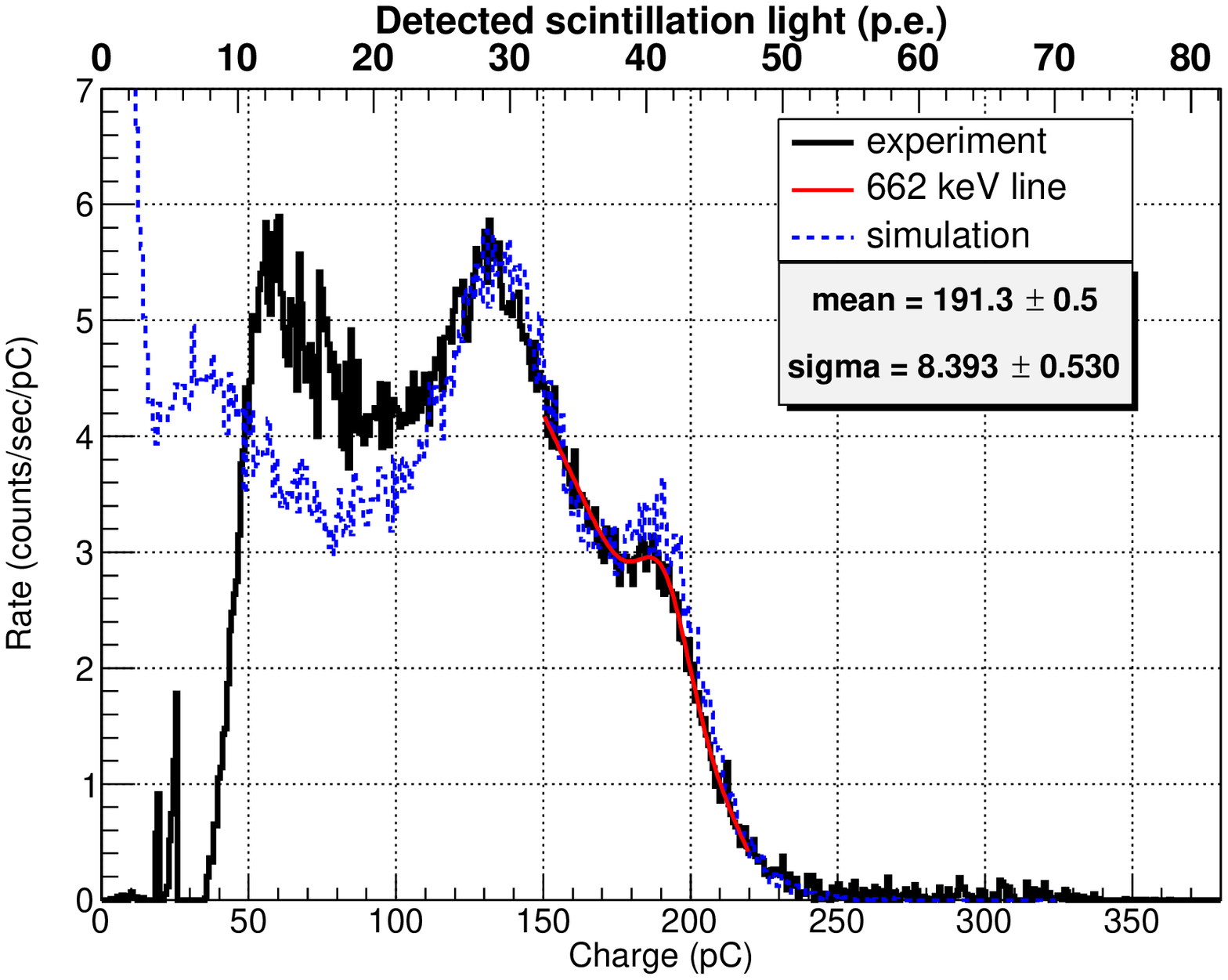}
\figcaption{\label{fig5}   Scintillation spectrum of $^{137}$Cs collimated at position B (solid black
histogram), as well as the Monte Carlo simulation result (dashed blue histogram). The full absorption
peak is fitted with the sum of a Gaussian and a linear function(red line).}
\end{center}
\begin{center}
\includegraphics[width=8.5cm]{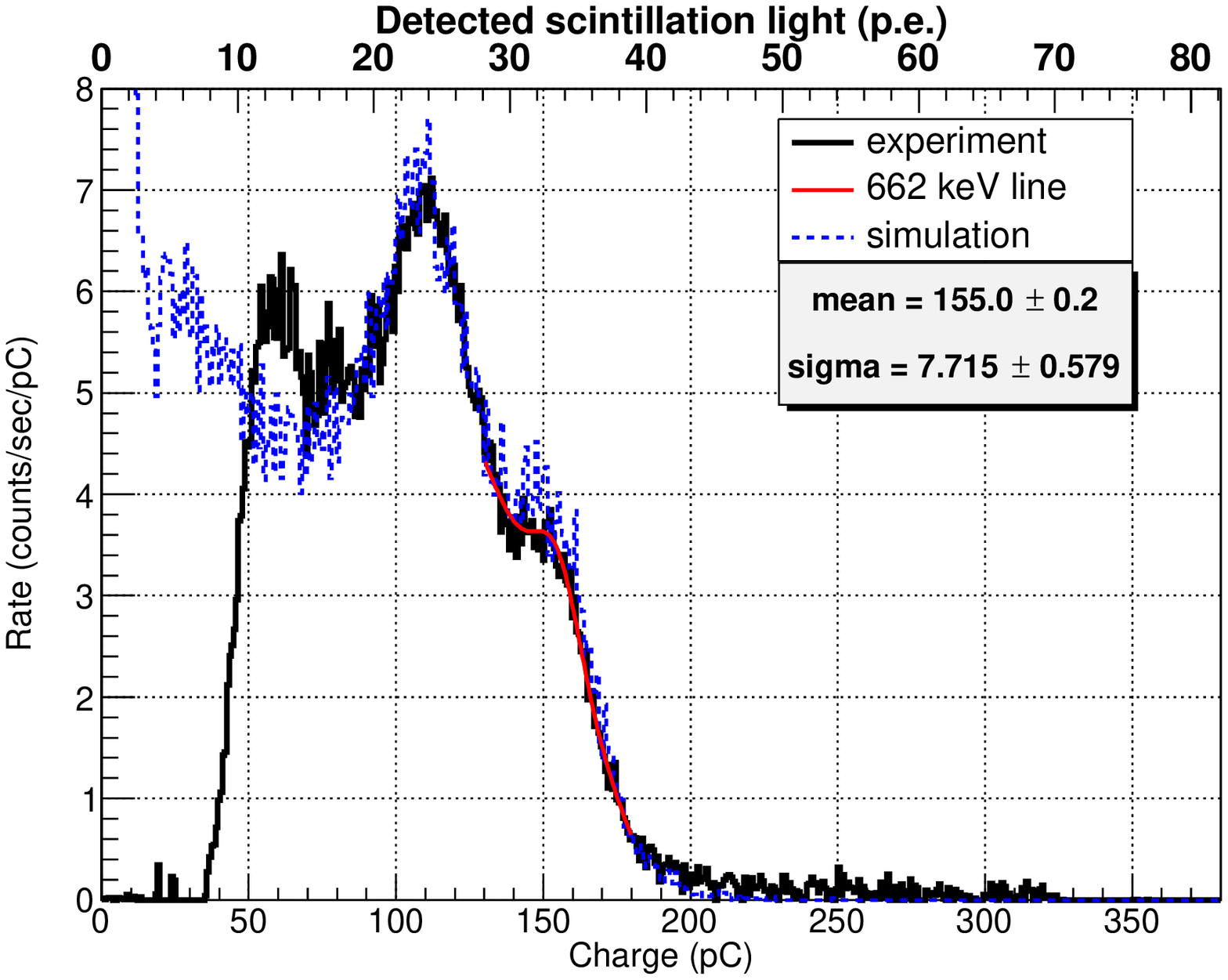}
\figcaption{\label{fig6}   Scintillation spectrum of $^{137}$Cs collimated at position C (solid black
histogram), as well as the Monte Carlo simulation result (dashed blue histogram). The full absorption
peak is fitted with the sum of a Gaussian and a linear function(red line).}
\end{center}

Each full absorption peak has been fitted with the sum of a Gaussian and a linear
function. The best-fit function, superimposed on the histogram, shows the mean value of
the full-absorption. According to the absolute gain of the PMT for single photoelectron of
4.625 pC, the photoelectron yield (${\rm PY}_\gamma$) for the 662 keV line, defined for each fit
as $\mu_p/E_\gamma$, can then be given (shown in Table~\ref{tab1}).
\begin{center}
\tabcaption{ \label{tab1}  photoelectron yield for each position of collimated Gamma source.}
\footnotesize
\begin{tabular*}{86mm}{ccccc}
\toprule
\shortstack{Gamma source\\position} & $\mu_p$/pC & p.e. & PY$_\gamma$/(p.e./keV) & Error/\% \\
\hline
A & 243.4 & 52.6 & 0.079 & 0.16 \\
B & 191.3 & 41.4 & 0.062 & 0.26 \\
C & 155.0 & 33.5 & 0.051 & 0.13 \\
\bottomrule
\end{tabular*}
\end{center}

We found that the photoelectron yield decreased with the increase of the distance from PMT window to
the position that the $^{137}$Cs source was collimated. It is certain that the distance from PMT
window to the location of the Gamma energy deposition significantly affected the number of
photons collected by the PMT. It is due to the depletion of the argon scintillation light
(via absorption mechanism) during transmission\cite{lab7,lab8}. Note that LAr is highly transparent
to visible light, so the depletion of visible light is negligible during transmission.
Meanwhile, because the reflectivity of Teflon is not strictly equal to 100\%, loss of photons occurs
in the process of reflection.

Some parameters of the prototype detector, such as the Surviving Fraction of the Ar$^*_2$ excimers ($S_{{\rm Ar}^*_2}$)
(defined as the ratio of the scintillation light yield in the experiment
to the yield recorded in the case of pure liquid Argon),
the absorption length for 128 nm photons in liquid argon ($L_{\rm abs}$) and the reflectivity of Teflon ($R_{\rm Teflon}$)
to UV and visible light (considered to be the same\cite{Teflon_Ref}), are affected by the LAr purity or Teflon polishing method and
therefore remain unknown in our experiment. By adjusting their values in the simulation, it is found that
the photoelectron yield is mainly determined by $S_{{\rm Ar}^*_2}$ while the ratio of the photoelectron yields
of different Gamma source positions is mainly determined by $L_{\rm abs}$ and $R_{\rm Teflon}$.
By ensuring that the simulation spectra are in good agreement with the experiment results, the parameters were
finally determined (listed in Table~\ref{tab2}).
\begin{center}
\tabcaption{ \label{tab2}  The values of the parameters determined in the simulation.}
\footnotesize
\begin{tabular*}{30mm}{cc}
\toprule
parameter & value \\
\hline
$S_{{\rm Ar}^*_2}$ & 4.0\% \\
$L_{\rm abs}$ & 1.0 m \\
$R_{\rm Teflon}$ & 84\% \\
\bottomrule
\end{tabular*}
\end{center}

The value of $S_{{\rm Ar}^*_2}$ (4\%) indicates that the contamination level of LAr was $\sim10^3$ ppm of
Nitrogen equivalent concentration according to the results of Ref.~\cite{lab8}. According to the previous
studies, Teflon, or PTFE,
has more than 90\% reflectivity for UV and visible light\cite{Teflon_Ref}. The Teflon reflectivity still has enough
space for improvement in our experiment. To obtain higher photoelectron and better detector resolution, it is
necessary to improve the LAr purity and the reflectivity of Teflon. In the next stage of experiment, the
purification system is proposed and new methods of Teflon polishing will be researched.

\subsection{WLS coating on the PMT window}

\label{WLS}
The PMT in the experiments above was coated with TPB by vacuum evaporation, with the thickness of
$\sim$0.3 mg/cm$^2$. The TPB coating shifts VUV light to visible light so that it can pass through
the PMT window and be detected by the PMT. But on the other hand the transparency of the PMT window
to visible light would be significantly reduced by the TPB coating. Therefore attention should be attached to
the TPB coating on the PMT window. A contrast experiment was carried out to investigate the influence
of the PMT window coating. In the contrast experiment, the PMT window was uncoated while other aspects
remain the same. The $^{137}$Cs source was collimated at position A and the spectrum is shown
in Fig.~\ref{fig7}.
\begin{center}
\includegraphics[width=9cm]{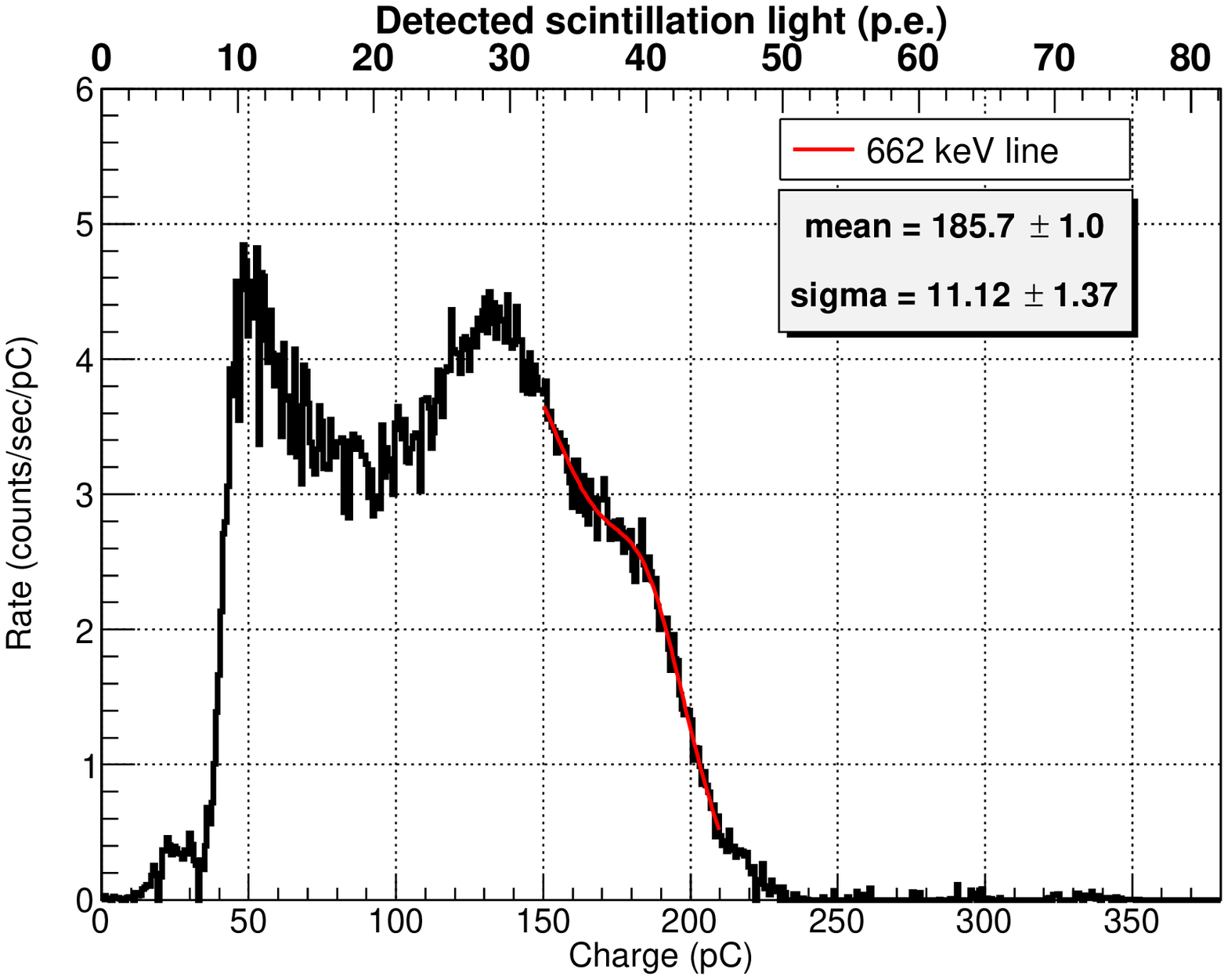}
\figcaption{\label{fig7}   Scintillation spectrum of $^{137}$Cs collimated at position A under
the condition that the PMT window was uncoated.}
\end{center}
The mean value of the full-absorption peak is 185.7 pC, which corresponds to about 40.2 photoelectrons.
The photoelectron yield can then be given as ${\rm PY}_\gamma = 0.061$ p.e./keV. We can draw a conclusion
that the photoelectron yield increases of about 30\%  with the presence of the PMT window coating. The result is
consistent with the one in Ref.~\cite{lab9}.

\section{Summary}

The China Dark Matter Experiment is designed to directly detect the dark matter with the high-purity
germanium detector. In the second phase CDEX-10, a LAr veto detector is employed to cool down and further
decrease the background of the HPGe detector. In order to develop
and optimize the light collection of the liquid-argon-based detector, a prototype was built and tested.

A gain calibration of the PMT has been done by using an LED calibration procedure. The absolute
gain of the PMT for single photoelectron is 4.625 pC under the liquid argon temperature.

We measured the photoelectron yield for 662 keV Gamma lines. The photoelectron yield turns out to be 0.079 p.e./keV
when the Gamma lines were collimated to position A. But the photoelectron yield decreased with
the increase of the distance from PMT window to the position that the $^{137}$Cs source was collimated.
It is probably due to the depletion of the argon scintillation light
(via absorption mechanism) during transmission and the photons loss in the process of reflection.
The good agreement between the experimental and simulation results has been reached and some important parameters
of the prototype detector have been determined by simulation.

A contrast experiment was carried out to investigate the influence of the PMT window coating. It shows
that PMT window coating improves the total light collection by about 30\%.

\section{Outlook}

As the veto detector of the low energy threshold HPGe detector in CDEX-10, LAr detector is expected
to have a good performance of energy threshold and veto efficiency which are mainly decided by the
photoelectron yield of the LAr detector. By reference to the LAr parameters obtained above, we can
simulate and predict the energy threshold and the veto efficiency of the CDEX-10 LAr veto detector under the
similar experimental conditions. Some preliminary results have been obtained.

To increase the photoelectron yield of the CDEX-10 LAr veto detector, the following methods will be considered:\\
1) Purifying LAr with a getter.\\
2) Optimizing the TPB coating process to improve the coating uniformity.\\
3) Researching a new polishing method of Teflon.\\
4) Replacing the PMTs with higher QE ones at reasonable cost.\\
5) Optimizing the layout of the PMTs to maximize the photocathode coverage.

\end{multicols}

\vspace{-1mm}
\centerline{\rule{80mm}{0.1pt}}
\vspace{2mm}

\begin{multicols}{2}

\end{multicols}

\clearpage

\end{CJK*}
\end{document}